\documentclass{article}

\usepackage{arxiv}

\usepackage[utf8]{inputenc} 
\usepackage[T1]{fontenc}    
\usepackage[colorlinks=true,linkcolor=blue,
            linkcolor = blue,
            urlcolor  = blue,
            citecolor = blue,]{hyperref}%
\usepackage{url}            
\usepackage{booktabs}       
\usepackage{amsfonts}       
\usepackage{float} 
\usepackage{amsmath}   
\usepackage{nicefrac}       
\usepackage{microtype}      
\usepackage{lipsum}		
\usepackage{graphicx}
\usepackage{natbib}
\usepackage{doi}

\title{Derivation of an Analytical Solution of a Forced   Cantilevered Tube Conveying  Fluid}

\author{ \href{https://orcid.org/0000-0003-4655-4351}{\includegraphics[scale=0.06]{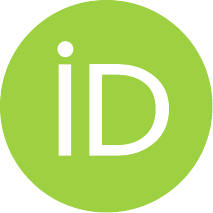}\hspace{1mm} Moussa ~Tembely}\thanks{moussa.tembely@concordia.ca} \\
	Department of Mechanical, Industrial \& Aerospace Engineering, \\
	 Concordia University\\
	Montreal, QC, Canada \\
}

\renewcommand{\shorttitle}{\textit{Analytical Solution of a Cantilevered Tube Conveying  Fluid}}

\hypersetup{
pdftitle={Derivation of an Analytical Solution of a Forced   Cantilevered Tube Conveying  Fluid},
pdfsubject={theory},
pdfauthor={Moussa~Tembely},
pdfkeywords={Tube conveying fluid; Forced vibration; Analytical solution; Fluid-structure interaction},
}

\begin{document}
\maketitle

\begin{abstract}
	In this paper, an analytical technique is proposed to obtain the forced response of a cantilevered tube conveying fluid. By considering the pipe subjected to an arbitrary harmonic force, either distributed or concentrated, an analytical solution is found using the Green's function method. The closed-form solution obtained satisfies the differential equations governing the vibrating tube conveying fluid. The proposed method, which provides exact solutions, is more accurate than the classical eigenfunction expansion or Galerkin's method and eliminates the need for eigenfunctions, eigenvalues, or infinite series.
\end{abstract}


\section{Introduction}
\label{Sec:Intro}

Pipes or tubes carrying fluid play a critical role in a wide range of engineering applications, including fuel pipelines, inkjet print-head nozzles, fuel-carrying tubes in space shuttle engines, heat exchangers, and piping systems in nuclear power plants. Understanding and controlling the forced vibrations of these tubes are essential for their efficient and safe operation \cite{Paidoussis2014,Tembely2011}. Additionally, feedback control methods used for beam vibrations can be adapted to mitigate unwanted vibrations in fluid-conveying tubes \cite{Lin1996}.

From a theoretical standpoint,  fluid--structure interactions involving moving boundary conditions remains one of the most  challenging problems in dynamical systems. The term ``moving boundary condition'' refers to situations where the boundary---such as the walls of a tube---is not stationary but moves or deforms in response to external forces acting on the fluid-conveying tube. This dynamic boundary significantly complicates the analysis of fluid--structure interactions.

For over 50 years these complexities have been the focus of extensive theoretical and experimental research. The challenges that have been explored include changes in the natural frequencies of the pipe system due to flowing fluid, vibrations induced by turbulent flows, and instabilities such as flutter (via a Hopf bifurcation) that occur when the flow velocity reaches a critical threshold \cite{Housner2021}. According to  \cite{Weaver1973}, the study of forced vibrations in pipes conveying fluid encompasses two main aspects: the physical aspect, which provides insights into the system's dynamics, and the analytical aspect, which involves developing techniques to predict the forced response of such systems. 
This work focuses on developing analytical techniques to predict the forced responses of pipes conveying fluid, contributing to a better understanding of these complex systems. Understanding their stability and vibrational characteristics is essential for safety and performance in practical applications. For instance, \cite{zhang_dynamical_2023} investigated the dynamic stability of cantilevered pipes under various lateral distributed loads using the differential quadrature method, finding that descending distributed loads significantly enhance system stability. \cite{Sepehri2020} examined the vibration of functionally graded damped beams with multiple fractionally damped absorbers, providing analytical and numerical solutions. Yang et al.~\cite{Yang2022} presented analytical solutions to peridynamic beam equations under static and dynamic loading, employing the central difference scheme for numerical solutions. \cite{Mojtaba2015} applied Green's function method to uniform Euler--Bernoulli beams at resonant conditions, introducing the Fredholm Alternative Theorem, but without considering fluid interactions or pointwise forces. Advancements in analytical solutions for pipe flow include \cite{Garcia2019,Garcia2020}, who developed analytical solutions for unsteady and transient turbulent pipe flow. As a simplified test case, they developed an analytical solution for starting turbulent pipe flow originated by a sudden pressure gradient increase, investigating the influence of Reynolds shear stress on velocity patterns in turbulent starting flows. \cite{Urbanowicz2023a,Urbanowicz2023b} reviewed Navier--Stokes solutions for accelerating pipe flow and advanced wall shear-stress modeling for water hammer phenomena. \cite{Henclik2018} analyzed an analytical solution for water hammer in pipelines with elastically attached valves. \cite{Bayle2023} proposed Laplace-domain fluid--structure interaction solutions for water hammer waves in a pipe. However, their approaches do not fully address the challenges associated with vibrating point forces in pipes conveying fluid.

 This paper presents an improved analytical strategy based on the Green's function method to develop closed-form solutions for a cantilevered tube conveying fluid, ensuring smoothness with continuous derivatives and satisfying the linear governing equations. The proposed method is applied under steady-flow conditions where the fluid velocity $U$, with a constant volumetric flow rate, is assumed to be constant. Additionally, the method does not require solving the free vibration problem to obtain eigenvalues and related eigenfunctions, which are typically necessary when using the classical Galerkin method.

\section{Tube Conveying  Fluid General Equation of Motion }
\begin{figure}[H]
    \centering

    \includegraphics[scale=0.4]{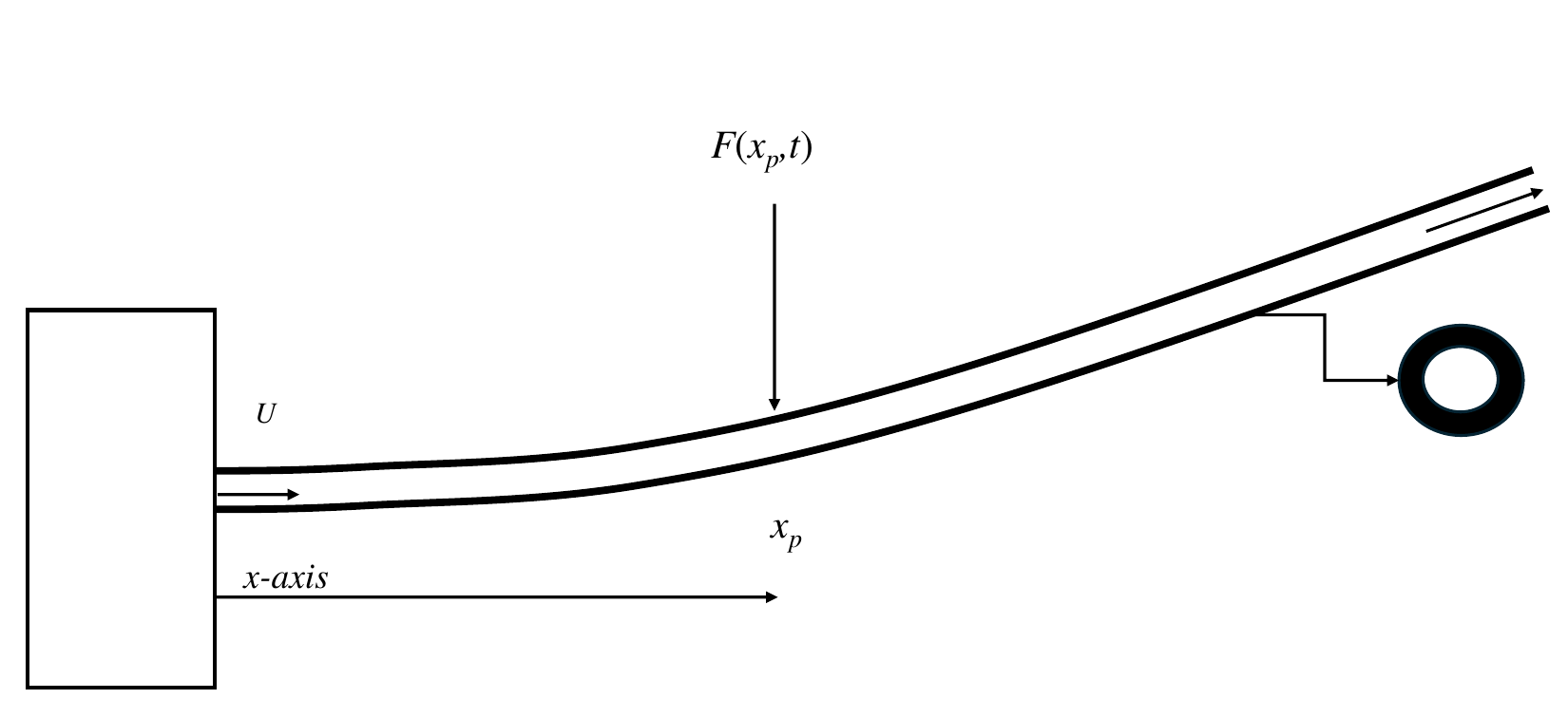}
    \caption{Schematic of a tube along the $x$-axis conveying fluid with velocity $U$, subjected to an external force $F(x, t)$.} 

    \label{Fig:tubefluid}
\end{figure}

The system under consideration is a straight cantilever tube fixed at one end on the $x$-axis. The fluid with an average velocity  $U$ enters the tube at the fixed end and exits at the free end. A force is applied to the tube, allowing it to move transversely, as depicted in Figure~\ref{Fig:tubefluid}. Furthermore, the force that induces tube motion is assumed to be harmonic. Unlike the unconstrained pipe dynamical stability studied by many others \cite{Weaver1973, Paidoussis2007}, the goal here is considerably different in the sense that we want to know the precise deflection of the tube, with the harmonic force considered either distributed or concentrated.

 \vspace{-3pt}

Neglecting the effects of dissipation and damping, the general equation of motion for a tube conveying fluid can be formulated based on the following assumptions: (i) the tube has a uniform annular cross-section; (ii) the length of the tube is significantly greater than its diameter; (iii) the effects of rotary inertia and shear deformation are neglected; \mbox{(iv) the }tube's centerline is inextensible; (v) the tube is elastic and initially straight; and \mbox{(vi) cross-sections} of the tube remain plane and perpendicular to the axis of the pipe containing a flowing liquid, in accordance with the Bernoulli--Euler beam theory. The equation governing the transverse displacement $v(x,t)$ can be expressed as follows \cite{Paidoussis2014, Ammari2000}:

\begin{equation} \label{eq-gene}
EI \frac{\partial^4 v}{\partial x^4} + M_fU^2 \frac{\partial^2 v}{\partial x^2} + 2UM_f \frac{\partial^2 v}{\partial t \partial x} + (m_t + M_f) \frac{\partial^2 v}{\partial t^2} = F(x, t) 
\end{equation}
where $EI$ is the flexural rigidity of the beam, $M_f$ is the mass per unit length of the fluid, $m_t$ is the mass per unit length of the beam, and $F(x, t)$ is the applied force as a function of position and time.

\section{Exact Analytical Solution}

We propose an approach based on Green's function method to solve Equation~(\ref{eq-gene}). Assuming the applied force is harmonic and localized, it can be represented as a point force:
\begin{equation}
F(x, t) = F_p \delta(x - x_p) \sin(\Omega_p t),
\end{equation}
where $F_p$ is the amplitude of the force, $\Omega_p$ is the frequency, and $\delta(x - x_p)$ represents a Dirac delta function centered at $x_p$, indicating that the force is applied at position $x_p$. For convenience, we reformulate the problem in a complex form by expressing the sinusoidal force as

\begin{equation} \label{eq;forcing}
F(x,t):=F_{p}\delta(x-x_{p})e^{j(\Omega_{p}t-\pi/2)}, \end{equation}
where $e^{i \Omega_p t}$ is the complex representation of the time-dependent oscillation. This allows us to work with simpler exponential functions, therefore simplifying the analysis of the system's harmonic response using Green's function.

Green's function technique is a well-known method for  solving  PDEs. A Green's function is the solution of the equation subject to a unit impulse. The technique can be applied as detailed below. We consider the general governing  equation
\begin{equation} \label{greeng}
\mathcal{L}u(\vec{x})=h(\vec{x}),\end{equation}
where $\mathcal{L}$ is a linear differential operator that involves derivatives with respect to spatial variables and possibly time. It acts on the unknown function $u(\vec{x})$, which depends on the spatial coordinates $\vec{x}$--which could represent ($x$,$y$,$z$) in 3D. The term $h(\vec{x})$ is a known function that represents the external forcing applied to the system. The objective is to determine $u(\vec{x})$ such that this equation is satisfied for the given $h(\vec{x})$.

We assume that the following equation, when solved with an impulse, yields Green's function \( G(\vec{x},\vec{x'}) \) as the solution:

\begin{equation} \label{greeng3}
\mathcal{L}G(\vec{x},\vec{x'}) = \delta(\vec{x} - \vec{x'}),
\end{equation}
where \(\delta(\vec{x} - \vec{x'})\) is the Dirac delta function. In addition,  we can express

\begin{equation}
h(\vec{x})=\int h(\vec{x}')\delta(\vec{x}-\vec{x}')d\vec{x'};\end{equation}
using  \eqref{greeng3}, and, with the linearity of \eqref{greeng}, one obtains 

\begin{equation}
u(\vec{x})=\int G(\vec{x},\vec{x'})h(\vec{x}')d\vec{x'}.\end{equation}
 
In the following, this result will be applied to solve the equation of motion of a forced tube conveying fluid with $\vec{x}$ simplified to $x$.

Based on the forcing term in \eqref{eq;forcing}, the solution of our problem is sought as 

\begin{equation} \label{solcplx}
w(x,t)=V(x)e^{j(\Omega_{p}t-\pi/2)},\end{equation}
where $V(x)$ is a function that describes the spatial variation of the solution. 
In addition,  the solution $v(x,t)$ of (\ref{eq-gene}) is the real part of $w(x,t)$. By substituting  \eqref{solcplx} into \eqref{eq-gene}, we obtain 

\begin{equation} \label{recast3} 
V''''(x) +a V''(x)+jbV'(x)-cV(x)=\frac{F_{p}\delta(x-x_{p})}{EI},
\end{equation} 
where  $a=M_f U^2/EI$, $b=2M_fU\Omega_p/EI$ and $c=(m_t+M_f)\Omega_p^2/EI$, the  solution of which is given by 

\begin{equation} \label{eq:solfsi}
V(x)=\frac{F_{p}}{EI}\intop_{0}^{L}\delta(x'-x_{p})G(x,x')dx'=\frac{F_{p}}{EI}G(x,x_p),\end{equation}
where we  determine the Green's function by solving the following equation:

\begin{equation} \label{recast3b} 
V''''(x) +a V''(x)+jbV'(x)-cV(x)=\delta(x-x'),
\end{equation} 
which is  performed by adopting the Laplace transform and its inverse. We then deduce 

\begin{equation} \label{eq:greenok}
\begin{array}{l}
 G(x,x')= \mbox{H} ( x- x' )  \sum\limits _{i=1}^4{\frac {{e^{{  \alpha_i} ( x-x' ) }}}{4{{ \alpha_i}}^{3}+2{  \alpha_i}{ a}+j{ b}}} \\[+1ex]
~~~~~~~~~~~~~~~~~~~~~~~~~~~+V ( 0 )  \sum\limits _{i=1}^4{\frac {{{ \alpha_i}}^{3}{e^{{ \alpha_i}x}}}{4{{ \alpha_i}}^{3}+2{ \alpha_i}{ a}+j{ b}}}+ V'(0)   \sum\limits _{i=1}^4{\frac {{{ \alpha_i}}^{2}{e^ {{ \alpha_i}x}}}{4{{ \alpha_i}}^{3}+2{ \alpha_i}{ a}+j{ b}}}\\[+1ex]
~~~~~~~~~~~~~~~~~~~~~~~~~~~~~~~+ \sum\limits _{i=1}^4{\frac {{e^{{ \alpha_i}x}}}{4{{\alpha_i}}^{3}+2{ \alpha_i}{ a}+j{ b}}} ( V'''  ( 0 ) + aV' ( 0 ) +  jbV ( 0 )  ) \\[+1ex]
~~~~~~~~~~~~~~~~~~~~~~~~~~~~~~~~~~~~~~~~~~~~~~~~~~~~~~~~~~~~~~~~+ \sum\limits _{i=1}^4{\frac {{ \alpha_i}{e^{{ \alpha_i}x}}}{4{{ \alpha_i}}^{3}+2 { \alpha_i}{ a}+j{ b}}} ( V'' ( 0 ) +aV ( 0)  ),\\
\end{array}
\end{equation}
where $H(x)$ is the Heaviside or step function,
    \begin{equation} 
\mbox{H} (x):= \begin{cases} 
  0  & \mbox {if } x< 0   \\
  1 & \mbox{if } x \ge 0, \end{cases}
  \end{equation} 
 and  $\alpha_i$, $ 1 \le i \le 4$  are the solutions of the following polynomial equation:
\begin{equation} 
Z^4 +a Z^2+jbZ-c=0.  \end{equation} 
This equation can be analytically solved using  Ferrari's method \cite{spiegel1968}.

Let  $\psi(x)$ be the function defined as

\begin{equation} 
 \psi(x)= \sum _{i=1}^4 \frac {e^{\alpha_i  x}}{4\alpha_i^{3}+2 \alpha_i a+j b}.
\end{equation} 

We then can write \eqref{eq:greenok} in the following form:

\begin{equation} \label{eq:greenok22}
\begin{array}{l}
 G(x,x') = \mbox{H} ( x- x' )  \psi(x-x')+  \left[ \psi'''(x) +a\psi'(x) 
+jb\psi(x) \right] V ( 0 )\\[+1ex]
~~~~~~~~~~~~~~~~~~~~~~~~~~~~~~~~~~~~~~~~~~~~~~~~~~+  [\psi''(x) +a\psi(x) ]V'(0)+\psi'(x)V''(0) +\psi(x)V'''(0).\\
\end{array}
 \end{equation}

The last step is now to decide on the different boundary conditions. For this purpose, we use the boundary conditions for the cantilever tube. At the fixed end (\(x = 0\)), the deflection \(v(x)\) is zero, i.e., \(v(0) = 0\), and the slope $\frac{dv(x)}{dx} =0 $. At the free end (\(x = L\)), both the bending moment \(EI \frac{d^2v(x)}{dx^2}\)  and the shear force \(EI \frac{d^3v(x)}{dx^3}\) are zero. Applying this to \eqref{eq:greenok22} leads to:
\begin{equation} \label{eqbc1}
V''(L)=0= \psi''(L-x')+ \psi'''(L)V''(0) +\psi''(L)V'''(0),
\end{equation} 
\begin{equation} \label{eqbc2}
V'''(L)=0= \psi'''(L-x')+ \psi^{(4)}(L)V''(0) +\psi'''(L)V'''(0),
\end{equation} 
where we use the fact that $0<x'\le L$ along the tube, thus leading to $\mbox{H}(L-x')=1$ and $\mbox{H}(0-x')=0$. 
 
 Solving the previous equations \eqref{eqbc1} and \eqref{eqbc2} leads to the determination of unknown boundary conditions:
\begin{equation} 
 V''(0)=\frac{\psi''(L)\psi'''(L-x')-\psi'''(L)\psi''(L-x')}{\psi'''(L)^2-\psi''(L)\psi^{(4)}(L)},  
\end{equation} 
\begin{equation} 
 V'''(0)=\frac{\psi^{(4)}(L)\psi''(L-x')-\psi'''(L)\psi'''(L-x')}{\psi'''(L)^2-\psi''(L)\psi^{(4)}(L)}.  
\end{equation}

%

Finally, we can  deduce the Green's function of a tube conveying fluid as

\begin{equation} \label{eq:greenok2}
\begin{array}{l}
 G(x,x') = \mbox{H} ( x- x' )  \psi(x-x')\\[+1ex]
~~~~~~~~~~~~~~~~~~~~~~~~~~~~~~~~~+ \frac{\psi''(L)\psi'''(L-x')-\psi'''(L)\psi''(L-x')}{\psi'''(L)^2-\psi''(L)\psi^{(4)}(L)}   \psi'(x)\\[+1ex]
~~~~~~~~~~~~~~~~~~~~~~~~~~~~~~~~~~~~~~~~~~~~~~~~~~~~~+ \frac{\psi^{(4)}(L)\psi''(L-x')-\psi'''(L)\psi'''(L-x')}{\psi'''(L)^2-\psi''(L)\psi^{(4)}(L)}    \psi(x).\\[+1ex]
\end{array}
 \end{equation}

Using \eqref{eq:solfsi}, we can finally establish the analytical solution of a tube conveying a fluid with a constant flow rate submitted to a punctual harmonic excitation as

\begin{equation} 
 v(x,t)=\Re [w(x,t)], 
\end{equation} 
where $w(x,t)$ is given by

\begin{equation} 
\begin{array}{l}
w(x,t) = \frac{F_p}{EI}\exp [j(\Omega_p t-\pi/2)] \{ \mbox{H} ( x- x_p )  \psi(x-x_p)\\[+1ex]
~~~~~~~~~~~~~~~~~~~~~~~~~~~~~~~~~+ \frac{\psi''(L)\psi'''(L-x_p)-\psi'''(L)\psi''(L-x_p)}{\psi'''(L)^2-\psi''(L)\psi^{(4)}(L)} \psi'(x)\\[+1ex]
~~~~~~~~~~~~~~~~~~~~~~~~~~~~~~~~~~~~~~~~~~~~~~~~~~+ \frac{\psi^{(4)}(L)\psi''(L-x_p)-\psi'''(L)\psi'''(L-x_p)}{\psi'''(L)^2-\psi''(L)\psi^{(4)}(L)}  \psi(x)  \}.\\
\end{array}
\end{equation}

\section{Model Validation}
We will validate the model by comparison with the series solutions, which are only valid for a tube without fluid, i.e., $U$~=~0. It is worth noting that the cantilevered beam subjected to a pointwise force can be solved analytically, and the summary of the solution (see Appendix \ref{appendixA}) is provided below, with the point force being represented as 
\begin{equation}
 F(x, t) = F_p \delta(x - x_p) \sin(\Omega_p t). 
\end{equation}

\textls[-15]{Using separation of variables, the analytical solution can be written as (see \mbox{Appendix \ref{appendixA}):}}

\begin{equation}
v(x, t) = \sum_{i=1}^{N} \frac{F_p V_i(x_p)}{m_t L (\omega_i^2 - \Omega_p^2)} V_i(x) \sin(\Omega_p t). \label{eq:solution}
\end{equation}

This represents the response of the empty tube (or cantilevered beam) in terms of its mode shapes (eigenfunctions) $(V_i(x)$ and natural frequencies $\omega_i$

\begin{equation}
V_i(x) = \cosh(\beta_i x) - \cos(\beta_i x) - \left( \frac{\sinh(\beta_i L) - \sin(\beta_i L)}{\cosh(\beta_i L) + \cos(\beta_i L)} \right) (\sinh(\beta_i x) - \sin(\beta_i x)), \label{eq:eigenfunctions}
\end{equation}
with
\begin{equation}
\omega_i = \beta_i^2 \sqrt{\frac{EI}{m_t}} \label{eq:natural_freq}
\end{equation}
obtained from the characteristic equation,

\begin{equation}
\cos(\beta_i L) \cosh(\beta_i L) + 1 = 0. \label{eq:char_eq}
\end{equation}

\begin{figure}[H]
    \centering
  \includegraphics[scale=0.5]{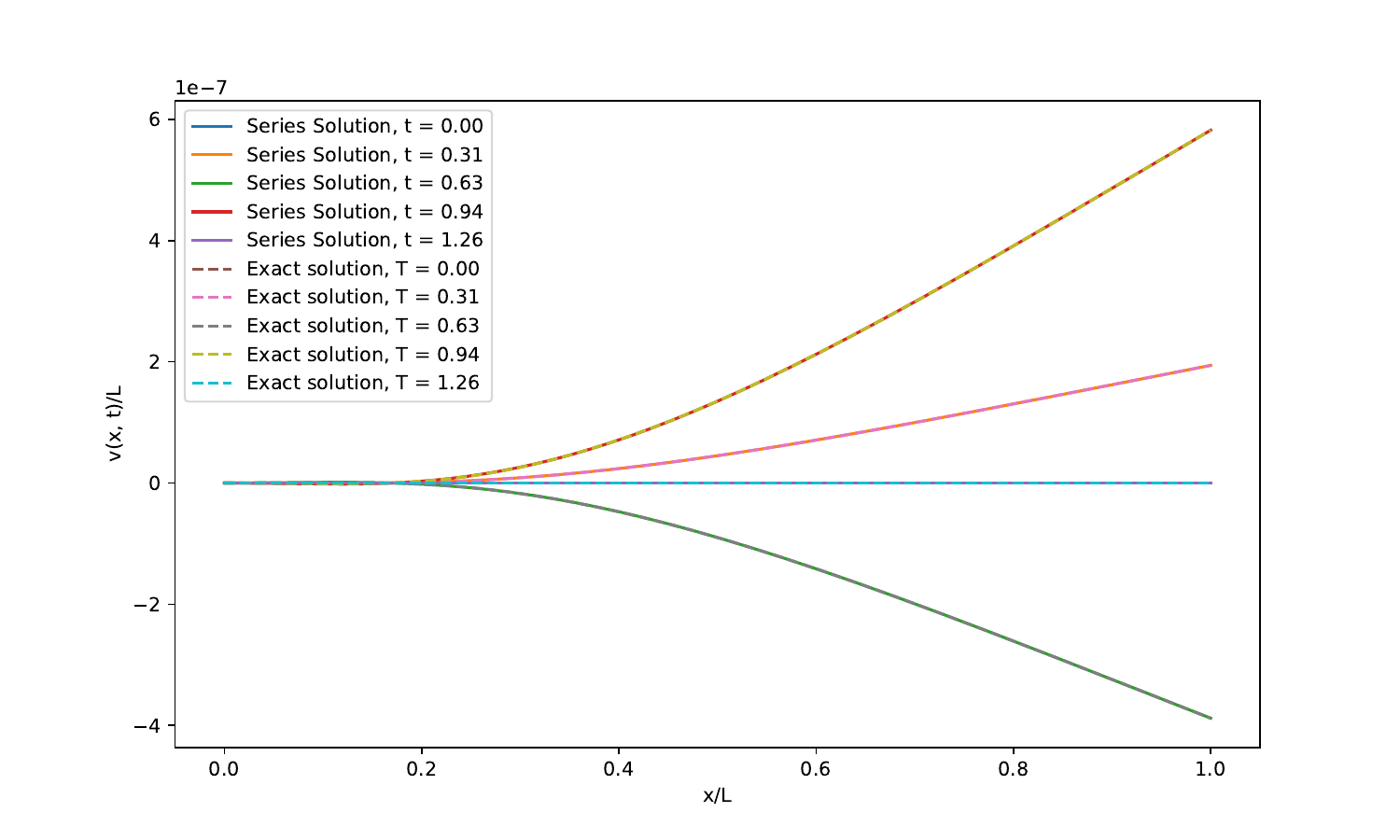}
    \caption{A comparison between the series and exact solution of a tube conveying fluid with \mbox{$L=10$ m,} $U$~=~0 m/s. }
    \label{Fig:2}
\end{figure}

\begin{figure}[H]
   \centering
   \includegraphics[scale=0.5]{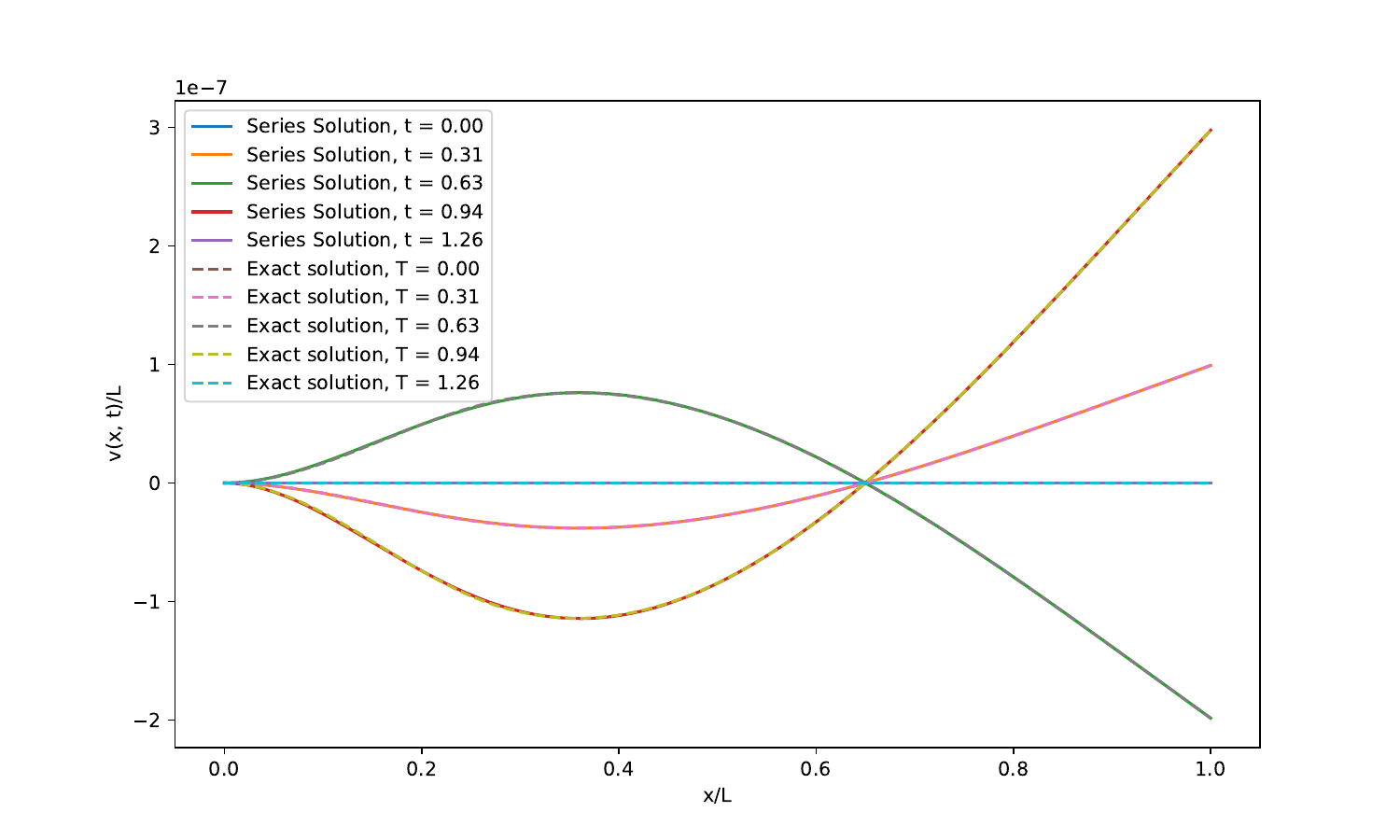}
    \caption{A comparison between the series and exact solution of an empty tube with $L=15$ m, \mbox{$U$~=~0 m/s.} }
    \label{Fig:3}
\end{figure}

\begin{figure}[H]
    \centering
   \includegraphics[scale=0.5] {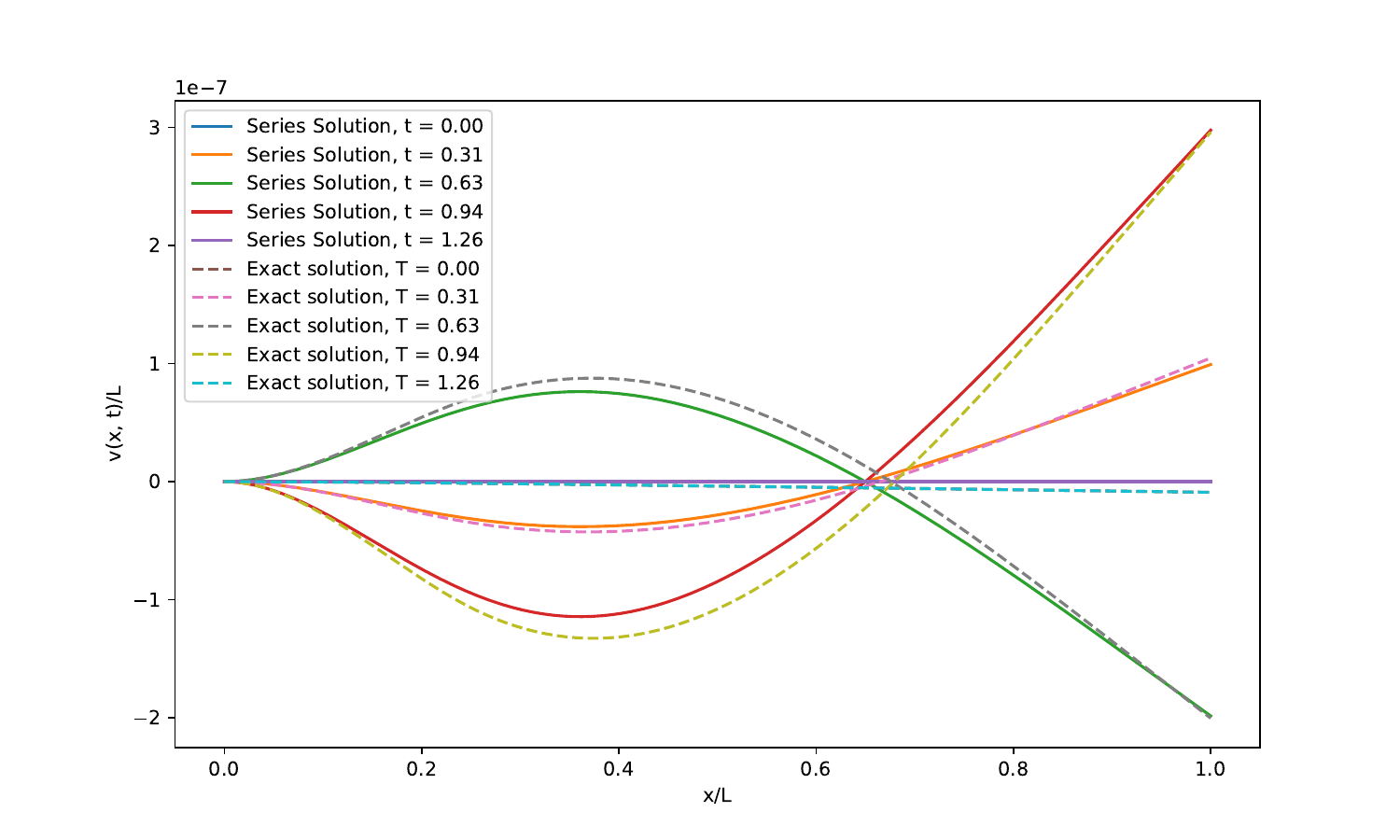}
    \caption{A comparison between the series and the exact solution of a tube conveying fluid with \mbox{$L=15$ m}, \mbox{$U$~=~0.1 m/s.} }
    \label{Fig:4}
\end{figure}

The results of the comparison between the series and closed-form Green's function solution are detailed  below. The properties of the tube and fluid are shown in  Table~\ref{Tab}. As we can see in Figures~\ref{Fig:2} and~\ref{Fig:3}  , a very good agreement is found for an empty tube, while the effect of the fluid can be observed in Figure~\ref{Fig:4} in terms of the tube's transient deflection. 

\begin{table}[h!] 
\centering 
\caption{Tube and fluid properties used for the model validation
}
\begin{tabular}{c c l c}
\hline
\textbf{Parameters} & \textbf{Value} & \textbf{Description} & \textbf{Unit} \\
\hline
$L$ & [1, 10, 15] & Length  & m \\
$\rho_t$ & 7800 & Density of the material & kg/m$^3$ \\
$\rho_f$ & [0, 1000] & Fluid density & kg/m$^3$ \\
$r_{\text{inner}}$ & 0.05 & tube inner radius & m \\
$r_{\text{outer}}$ & 0.06 & tube outer radius & m \\
$E$ & $2 \times 10^{11}$ & Young's modulus & Pa \\
$U$ & [0, 0.1, 1] & velocity & m/s \\
$\Omega_p$ & 10.0 & Excitation frequency & rad/s \\
$F_p$ & 10.0 & Amplitude of the forcing function & N \\
$x_p$ & $L$/4 & Position where the force is applied & m \\
\hline
$I$ & $I = \frac{\pi}{4} (r_{\text{outer}}^4 - r_{\text{inner}}^4)$ & Area moment of inertia &  -\\
$m_t$ & $m_t = \rho_t \pi (r_{\text{outer}}^2 - r_{\text{inner}}^2)$ & Mass per unit length of the tube & - \\
$M_f$ & $M_f = \rho_f \pi r_{\text{inner}}^2$ & Added mass per unit length due to fluid & - \\
\hline
$a$ & $a = \frac{M_f U^2}{EI}$ & - &  \\
$b$ & $b = \frac{M_f U \Omega_p}{EI}$ & - & -  \\
$c$ & $c = \frac{(m_t + M_f) \Omega_p^2}{EI}$ & - & - \\

\hline
\end{tabular}

\label{Tab}
\end{table}

\begin{figure}[H]
   \centering
   \hspace{-1.1cm} \includegraphics[scale=0.5]{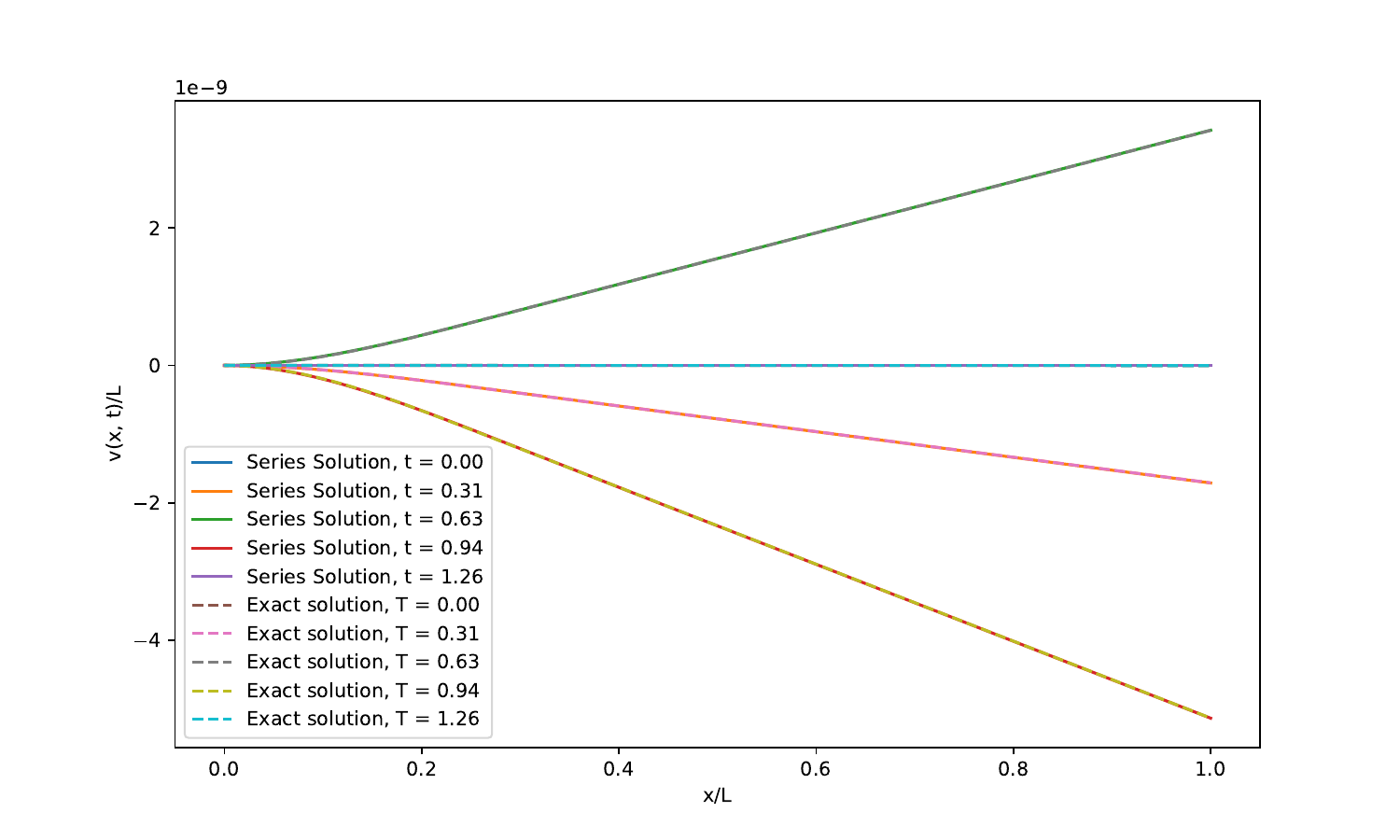}
    \caption{A comparison between the series and exact solution of a tube conveying fluid with \mbox{$L=1$ m,} $U$~=~1 m/s. }
    \label{Fig:5}
\end{figure}

However, the fluid velocity does not affect the pipe deflection, as shown below when the length of the tube is taken as $L$~=~1 m (Figure~\ref{Fig:5}).
 
 \section{Conclusions}
 Considering a general equation of motion for a tube-carrying fluid subjected to harmonic excitation, an analytical technique based on the Green's function method has been developed to find a solution for a tube conveying fluid. The derived exact solution is more accurate than the conventional series expansion or Galerkin's method, and it does not require knowledge of the free vibration problem's eigenfunctions or eigenvalues. The proposed method might also be extended to include moment solicitation and other configurations, such as multi-span tubes, different boundary conditions, and structural vibrations.


\bibliographystyle{unsrtnat}
\bibliography{Biblio}  

\appendix 
\section{Cantilevered Tube Series Solutions for Pointwise Harmonic Forcing } \label{appendixA}
\counterwithin*{equation}{section} 
\renewcommand\theequation{\thesection\arabic{equation}} 
\setcounter{equation}{0}    

We derive here a series solution for a vibrating cantilevered tube subjected to a harmonic pointwise force using the equation of motion for a vibrating beam:
\begin{equation} \label{GrindEQ__27_} 
EI\frac{\partial ^{4} v}{\partial x^{4} } +m_{t} \frac{\partial ^{2} v}{\partial t^{2} } =F(x,t)=F_p \delta (x-x_{p} ) \sin(\Omega_p t),
\end{equation} 
with $m_{t} =\rho _{t} S_{a} $, $S_{a} $ being the annular area of the tube  and $\rho_t$ being the tube density. 
  $F(x,t)$ represents a pointwise force per unit length applied at a specific point $x_p$ on the tube.
On the one hand, the boundary conditions for the tube displacement, along with the cantilevered condition at $x = 0$, are as follows:

\begin{equation}
v(0,t)=\frac{\partial v}{\partial x} (0,t)=0.\end{equation} 

 On the other hand, the bending moment $M=EI \kappa= \dfrac {EI \partial^2 v}{\partial x^2}$ and  the shear force $Q=-\dfrac{EI \partial^3 v}{\partial x^3}$ vanish at $x=L$, leading to :
\begin{equation}
\frac{\partial ^{2} v}{\partial x^{2} } (L,t)=\frac{\partial ^{3} v}{\partial x^{3}} (L,t)=0. \end{equation} 

For our problem, the initial conditions for deflection and velocity are both zero
 
\begin{equation}
v(x,0)=\frac{\partial  v}{\partial x } (x,0)=0. \end{equation} 

We adopt  the classical separation of variables method to solve Equation \eqref{GrindEQ__27_}   by assuming  that the displacement can be split  into two parts: the first one depends on position and the second one on time.  Then  series decomposition or Galerkin's  method is used to determine unsteady solution decomposed in orthonormalized mode as

\begin{equation} \label{GrindEQ__29_} 
v(x,t)=\sum _{i=1}^{N}q_{i} (t)V_{i}  (x), 
\end{equation} 
where the $V_{i} $  are the eigenfunctions or modal shapes determined thanks to free vibration equation and obtained as follows. We set

\begin{equation} 
 v(x,t)=V(x)T(t),
\end{equation} 

and substituting  this expression  into the homogeneous form of Equation \eqref{GrindEQ__27_}  results in:

\begin{equation} \label{GrindEQ30a} 
-\frac{EI V''''}{m_t V } (x)=\frac{\ddot{T}(t)} {T(t)} =-\omega^2. 
\end{equation}

The modal shapes could be determined by solving 

\begin{equation} \label{GrindEQ30b} 
\frac{d^{4} V}{dx^{4} } (x)-\beta^4 V=0, 
\end{equation} 
where  $\beta ^{4} =\dfrac{m_{t} \omega ^{2} }{EI}$, with the boundary conditions  given by

\begin{equation} \label{bcv}
  V(0)=V'(0)=0 ~  \mbox {and}~   V''(L)=V'''(L)=0. \end{equation} 

The solution to Equation \eqref{GrindEQ30b} is
\begin{equation} \label{bcv2}
V(x)=A\sin (\beta x)+B \cos (\beta x)+C\sinh  (\beta x)+D\cosh (\beta x). \end{equation} 

Using the boundary conditions from Equation \eqref{bcv} in this solution Equation \eqref{bcv2}  imposes  constraints to eliminate the trivial solution, 

\begin{equation}
\det \left[ \begin{matrix}
   0 & 1 & 0 & 1  \\
   1 & 0 & 1 & 0  \\
   -\sin \beta L & -\cos \beta L & \sinh \beta L & \cosh \beta L  \\
   -\cos \beta L & \sin \beta L & \cosh \beta L & \sinh \beta L  \\
\end{matrix} \right]=0,
\end{equation} 
 which leads to the following classical transcendental  or characteristic equation,

\begin{equation} \label{GrindEQ__32_} 
\cos (\beta L)\cosh (\beta L)+1=0. 
\end{equation} 

The solution of Equation \eqref{GrindEQ__32_}  could be approximated using an iterative method with an initial guess value of 

\begin{equation}
 \beta_{i} ^{(0)} L=(2i-1)\frac{\pi }{2}, 
\end{equation} 

 Otherwise,  a more accurate asymptotic approximation could be found as follows by setting: 

\begin{equation}
B_i =(2i-1)\frac{\pi }{2} +\varepsilon _{i}. 
\end{equation} 
\textls[-15]{using Equation \eqref{GrindEQ__32_}, we compute the correction $\varepsilon _{i} $ to find the following good approximation}, 

\begin{equation}
\beta _{i} L\approx B_i=(2i-1)\frac{\pi }{2} +\frac{(-1)^{i-1} }{\cosh \left\{(2i-1)\frac{\pi }{2} \right\}}.
\end{equation} 

 Finally,  the eigenfunctions  $V_{i} (x)$ could be expressed  after some mathematical manipulation using Equations \eqref{GrindEQ__32_} and \eqref{bcv2} as 
\begin{equation} \label{GrindEQ__33_} 
V_{i} (x)=\cosh (\beta _{i} x)-\cos (\beta _{i} x)-\frac{\sinh (\beta _{i} L )-\sin (\beta _{i} L )}{\cosh (\beta _{i}L )+\cos (\beta _{i}L )} (\sinh (\beta _{i} x)-\sin (\beta _{i} x)),
\end{equation} 
with $\omega_i =\beta_i ^{2} \sqrt{\dfrac{EI}{m_{t} } }$ the  eigenfrequencies.

The orthogonality of modal  shapes leads to:
\begin{equation} \label{eq:orthom}
  \int _{0}^{L}V_i V_j dx=L \delta_{ij}, 
\end{equation} 
 with  $\delta_{ij}$ the Kronecker delta ( 1 for $ i=j$ and  0 for $ i \ne j$ ).

Finally, the transversal motion of the tube using an eigenfunction  expansion can be expressed in the series solution as 

\begin{equation} \label{eq:solok0} 
v(x,t)=\sum _{i=1}^{N}  \tilde{c}_i V_{i}  (x) \sin(\Omega_p t),
\end{equation} 
where we take the temporal part of the form $q_i(t)=\sin(\Omega_p t)$.

 Substituting Equation \eqref{eq:solok0} in Equation \eqref{GrindEQ__27_}, multiplying by $V_{j} $,  and integrating along the tube from 0 to $L$, one finds, using modal shape orthogonality properties, Equation \eqref{eq:orthom}:

\begin{equation} \label{eq:ci} 
 \tilde{c}_i=\frac{F_p V_i(x_p)}{m_t L (\omega_i^2 - \Omega_p^2)}. 
\end{equation}

\end{document}